# A Novel Method for Fundamental Interaction Studies with Electrostatic Ion Beam Trap

S. Vaintraub[1,2], M. Hass[2], O. Aviv[2], O. Heber[2], I. Mardor[1]

[1] *Soreq NRC, Yavne 81800, Israel*
[2] *Weizmann Institute of Science, Rehovot 76100, Israel*

**INTRODUCTION**

Trapped radioactive atoms present exciting opportunities for the study of fundamental interactions and symmetries. For example, detecting beta decay in a trap can probe the minute experimental signal that originates from possible tensor or scalar terms in the weak interaction. Such scalar or tensor terms affect, e.g., the angular correlation between a neutrino and an electron in the beta-decay process, thus probing new physics of "beyond-the-standard-model" nature [1].

In particular, this article focuses on a novel use of an innovative ion trapping device, the Electrostatic Ion Beam Trap (EIBT) [2]. Such a trap has not been previously considered for Fundamental Interaction studies and exhibits potentially very significant advantages over other schemes. These advantages include improved injection efficiency of the radionuclide under study, an extended field-free region, ion-beam kinematics for better efficiency and ease-of-operation and the potential for a much larger solid angle for the electron and recoiling atom counters.

In the following we briefly present the theory of beta decay formalism, with special emphasis on the $^6$He beta decay. The subsequent sections include a brief method description and preliminary results. We conclude with a brief summary.

**THEORY**

The β-decay transition rate W (inverse lifetime) in case of non-oriented nucleus is given by[3]

$$dW \propto \xi\left(1 + a\frac{\vec{p}_e \cdot \vec{p}_\nu}{E_e E_\nu} + b\frac{m_e}{E_e} + ...\right) \propto \xi\left(1 + \frac{p_e}{E_e}a \cdot \cos\theta_{e\nu} + ...\right)$$

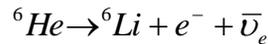
Beta-neutrino correlation coefficient

$a$, $b$ and others are the beta decay coefficients (more coefficients also exists in case of polarized nuclei[3]). In our recent study we concentrate on the beta-neutrino correlation coefficient - $a$, which can be measured completely unbiased in case of $^6$He beta decay, as will be shown below. The following analysis can be very similarly attuned to others beta decay coefficients and radio-nuclei.

The $^6$He nuclide has a half-life of 807 ms and is decaying into $^6$Li, electron and electron anti-neutrino

$$^6He \rightarrow {}^6Li + e^- + \bar{\upsilon}_e$$

The most general way to write the beta decay Hamiltonian[3] is:

$$H_\beta^{^6He} = \sum_{i=S,P,V,A,T} \overline{(^6Li)}\hat{O}_i(^6He)[\bar{e}\hat{O}_i(C_i + C_i'\gamma_5)\upsilon_e] + h.c.$$

where S,P,V,A and T stands for Scalar, Pseudo-scalar, Vector, Axial-vector and Tensor respectively. Then, the beta-neutrino correlation coefficient then can be expressed by the Gamow-Teller (GT) and Fermi (F) matrix elements along with eight coupling constants C, C' [3]:

$$a\xi = |F|^2 \left[|C_V|^2 - |C_S|^2 + |C'_V|^2 - |C'_S|^2\right] + \frac{|GT|^2}{3}\left[|C_T|^2 - |C_A|^2 + |C'_T|^2 - |C'_A|^2\right]$$

and

$$\xi = |F|^2 \left[|C_V|^2 + |C_S|^2 + |C'_V|^2 + |C'_S|^2\right] + |GT|^2 \left[|C_T|^2 + |C_A|^2 + |C'_T|^2 + |C'_A|^2\right]$$

The $^6$He beta decay is a $J^\pi=0^+$ to $J^\pi =1^+$ pure Gamow-Teller decay. In that case the beta-neutrino correlation coefficient can be expressed only by Axial and Tensor coupling constants in the Hamiltonian, since the Fermi matrix element is zero. In the Standard Model (SM) only the V-A interactions can occur, therefore only the axial coupling constants remain[3]:

$$|C_T|^2 = |C'_T|^2 = 0 \quad |C_A|^2 = |C'_A|^2 = 1$$

Consequently the transition rate for pure Gamow-Teller decays in the SM frame has the following simple form,

$$dW_{SM}(\theta_{ev}) \propto \left(1 - \frac{p_e}{3E_e}\cos\theta_{ev}\right)$$

where the squared GT matrix element can be brought out of brackets, hence its absolute value is not important. On the contrary, in a general case, beta decay may be neither a pure Gamow-Teller nor a pure Fermi decay. As a result any beta decay coefficient in the transition rate equation would be matrix elements depended, thus nuclear structure theory depended.

**METHOD**

The main purpose of the experiment is a **high precision measurement** of the beta decay coefficients. In the following we present a novel approach – that of using an Electrostatic Ion Beam Trap (EIBT). Following the pioneering work at the WI[2], the EIBT is being used worldwide (Germany, France, India, Japan and USA) in atomic and molecular physics experiments[4]. However, injecting and trapping radioactive nuclei for Fundamental Interaction studies in an EIBT device is an entirely novel direction that has never been tried before. This direction offers significant improvements in detection efficiency (paramount consideration when working with radioactive beams of limited intensity) and ease-of-of-operation.

The EIBT was developed in Weizmann Institute of Science in 1996[2] for storing ions at typical energy of few keV. The principle of operation of an EIBT[5] is based on the analogy to classical optical resonator: it is possible to store a beam of photons between two spherical mirrors separated by distance *L* if the focal length *f* fulfils the stability condition, namely: L/4 < f < ∞. A simple scheme of the EIBT setup is shown in Figure 1. The basic EIBT consist of a set of eight electrodes both acting as an electrostatic mirror by producing a retarding field which reflects the beam along its path and focuses it on the lateral direction. Thus the ions bounce back and forth between the two mirrors. The typical distance between the electrostatic mirrors is 500 mm.

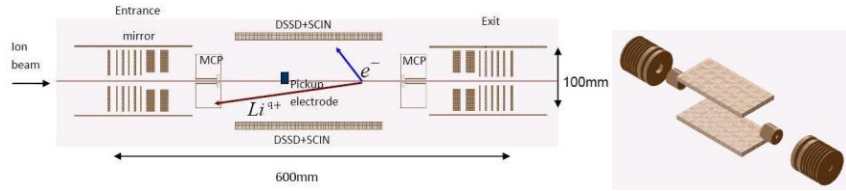

Figure 1. *A schematic view of the ESBT for β-decay studies. The radioactive ion, like $^6$He, moves with $E_k$~10 keV between the reflecting electrodes. The β electrons are detected in position sensitive counters while the recoiling ions, due to kinematical focusing, are detected with very high efficiency in either one (determined by the instantaneous direction) of the annular MCP counters.*

The trap assembly consists of two mirrors which reflect the beam and focus it on the lateral axis (see Fig. 1). Ions with kinetic energy of ~10 keV are injected through the grounded entrance mirror, when the ions fill the trap; potentials on the entrance mirror are quickly raised (<50ns) so that the ions oscillate back and forth between the mirrors. For $^6$He$^+$, the revolution time is ~4 microseconds. A pickup electrode is located in the trap center which is used to continuously monitor the bunch location. Therefore, position of the decay is known. The products of the decay (recoil Li nuclei and electron) are detected by a set of dedicated detectors. Two annular position sensitive Multi-Channel-Plate (MCP) detectors are situated along the main trap axis with the purpose of counting the recoil Li. These MCPs have central holes which allow the stable trajectories of the stored ions (such technology was demonstrated before[6]). The electrons are counted by two detectors located above and below the optical axis. The detectors are composed of plastic scintillators coupled to double sided silicon strip detectors (DSSD), thereby obtaining both energy and position of the beta particle.

**PRELIMINARY SUMULATION RESULTS**

Simulations are crucial to demonstrate the concept validity and, later on in the project itself, to optimize both the bunch dynamics and the detection geometry.
The construction of the dedicated set-up of an EBIT for beta-decay studies should naturally follow such simulations and take advantage of the gained insight. This process is currently ongoing and we present below first and preliminary results. The observable parameters (recoil time of flight, position of the ion bunch, position of particles on detectors, electron energy), allow the full reconstruction of the momentum of the undetected neutrino particle in the beta decay. Thus the angular correlation between electron and the neutrino upon decay can be studied.
The accuracy of the neutrino momentum is mainly limited by the certainty of the position of the decay, which can occur anywhere inside the ion bunch. While only the location of the bunch is measured. Previous works on the subject show that under special trapping conditions ("self bunching") [7], the ion bunch size can be reduced to few cm. We plan to investigate this issue further in the hope of achieving bunch size of typically 1 cm. Moreover, a study of the influence of the bunch size (e.g. the uncertainty of the decay within the bunch) on the sensitivity of our measurement is now underway and preliminary results are shown in Fig. 2 for different bunch sizes. As we can see small bunch sizes (~<2 cm) are almost identical to a point beta decay occurrences.

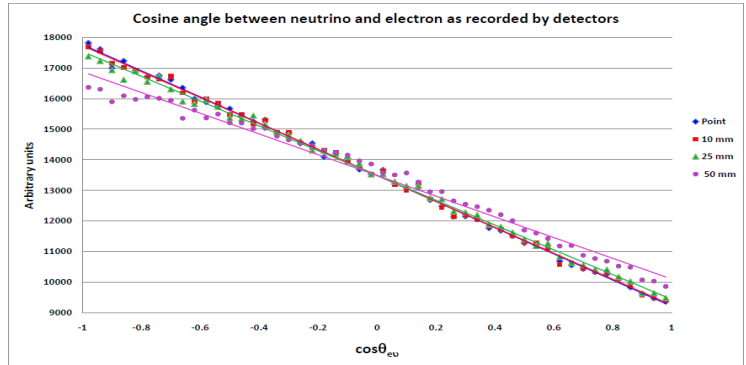

Figure 2. *GEANT4*[8] *simulation of beta-neutrino correlation with different bunch sizes.*

**SUMMARY AND FUTURE PLANS**

This research direction is also closely related in turn to the intensive R&D efforts that are currently under way in our laboratory to facilitate the production and extraction of such radioactive nuclei ($^6$He and $^8$Li, for example) in record yields, using neutron-induced reactions[9]. We are planning to use an intense d+t commercial neutron generator that provides ~14 MeV neutrons at a fluxes reaching $10^9$ - $10^{10}$ n/s. The purchase of such a generator has recently been approved by the WI. At a later stage, we envisage taking advantage of uniquely high yields of radioactive nuclei that will become available at the new and modern SARAF accelerator at Soreq.

Though the research program is only at the beginning, it seems robust and promising in the light of the recent calculations and our growing understanding of the system.